\documentclass[pdflatex,sn-mathphys-num]{sn-jnl}

\usepackage{graphicx}%
\usepackage{multirow}%
\usepackage{amsmath,amssymb,amsfonts}%
\usepackage{amsthm}%
\usepackage{mathrsfs}%
\usepackage[title]{appendix}%
\usepackage{xcolor}%
\usepackage{textcomp}%
\usepackage{manyfoot}%
\usepackage{booktabs}%
\usepackage{algorithm}%
\usepackage{algorithmicx}%
\usepackage{algpseudocode}%
\usepackage{listings}%

\usepackage{float}

\theoremstyle{thmstyleone}%
\theoremstyle{thmstyletwo}%

\theoremstyle{thmstylethree}%

\begin{document}

\title[Article Title]{Meta-operators for all-optical image processing}

\author[1]{\fnm{Linzhi} \sur{Yu}}\email{linzhi.yu@tuni.fi}

\author[1]{\fnm{Haobijam J.} \sur{Singh}}\email{johnson.singh@tuni.fi}

\author[1]{\fnm{Jesse} \sur{Pietila}}\email{jesse.pietila@tuni.fi}

\author*[1,2]{\fnm{Humeyra} \sur{Caglayan}}\email{h.caglayan@tue.nl}

\affil*[1]{\orgdiv{Department of Physics}, \orgname{Tampere University}, \city{Tampere}, \postcode{33720}, \country{Finland}}

\affil[2]{\orgdiv{Department of Electrical Engineering and Eindhoven Hendrik Casimir Institute}, \orgname{Eindhoven University of Technology}, \city{Eindhoven}, \postcode{5600 MB}, \country{The Netherlands}}

\abstract{All-optical image processing offers a high-speed, energy-efficient alternative to conventional electronic systems by leveraging the wave nature of light for parallel computation. However, traditional optical processors rely on bulky components, limiting scalability and integration. Here, we demonstrate a compact metasurface-based platform for analog optical computing. By employing double-phase encoding and polarization multiplexing, our approach enables arbitrary image transformations within a single passive nanophotonic device, eliminating the need for complex optical setups or digital post-processing. We experimentally showcase key computational operations, including first-order differentiation, cross-correlation, vertex detection, and Laplacian differentiation. Additionally, we extend this framework to high-resolution 3D holography, achieving subwavelength-scale volumetric wavefront control for depth-resolved reconstructions with high fidelity. Our results establish a scalable and versatile approach to computational optics, with applications including real-time image processing, energy-efficient computing, biomedical imaging, high-fidelity holographic displays, and optical data storage, driving the advancement of intelligent optical processors.}


\maketitle

\section{Introduction}\label{sec1}

\noindent The increasing reliance on image-driven applications, from deep learning to real-time analytics, demands high-speed, energy-efficient processing. However, conventional digital architectures, which rely on electronic circuits for image transformations, face fundamental limitations due to analog-to-digital conversion latency, sequential data handling, and power dissipation constraints. As transistor scaling nears physical limits, thermal dissipation becomes a major bottleneck, restricting further performance gains~\cite{williams2017s}. These challenges have motivated the exploration of alternative computing paradigms capable of real-time, low-power image processing. Optical computing overcomes these limitations by enabling parallel information processing without analog-to-digital conversion latency. Unlike electronic processors, which sequentially process pixel data, optical systems operate on spatially encoded information in a fully passive manner. This enables ultrafast, energy-efficient computation while reducing the processing load on digital systems. As a result, optical computing provides a promising pathway toward hybrid optical-digital architectures for high-speed image processing~\cite{solli2015analog,minzioni2019roadmap,mcmahon2023physics,hu2024diffractive}.

Although optical image processing has been explored for decades~\cite{javidi1994real,he2022computing}, its practical adoption has been limited by the bulky footprint of conventional optical components. Metasurfaces—artificial planar optical elements composed of subwavelength resonators (meta-atoms)—have emerged as a compact and highly versatile alternative for light-field manipulation~\cite{yu2011light,lin2014dielectric,kuznetsov2024roadmap}. Their ability to precisely control phase, amplitude, and polarization at subwavelength scales has led to the miniaturization of optical systems, including telescopes~\cite{zhang2022high,park2024all}, spectrometers~\cite{faraji2018compact,wen2024metasurface}, and interferometers~\cite{zhou2019optical,zhou2021two,yu2025multifunctional}. Beyond these applications, metasurfaces are increasingly explored for computational optics, enabling compact optical systems for analog computing and signal processing tasks~\cite{silva2014performing,kwon2018nonlocal,guo2018photonic,cordaro2023solving}.

Various metasurface-based approaches have been proposed for image processing, including image differentiation~\cite{zhou2019optical,zhou2021two,huo2020photonic,kim2022spiral,zhou2020flat,cotrufo2024reconfigurable,zhou2024laplace,wang2022single,fu2022ultracompact,tanriover2023metasurface,swartz2024broadband,deng2024broadband}, object detection~\cite{lin2018all,wang2022single,zheng2022meta,zheng2024multichannel}, image denoising~\cite{icsil2024all}, and coordinate transformations~\cite{sun2017transformation,ding2020metasurface,zhang2023all}. However, most existing studies focus on single-function implementations~\cite{zhou2019optical,zhou2021two,huo2020photonic,zhou2020flat,kim2022spiral,fu2022ultracompact,swartz2024broadband,tanriover2023metasurface,zhang2023all,deng2024broadband,zhou2024laplace,cotrufo2024reconfigurable} or require bulky optical setups that operate only at long wavelengths~\cite{lin2018all,wang2022single,icsil2024all,swartz2024broadband}, limiting their applicability for general-purpose image processing. Such constraints are particularly problematic in the visible range, which remains the most widely used spectral region for imaging due to its compatibility with human vision, high-resolution optical systems, and widespread applications in cameras, displays, and microscopy.

\begin{figure}[H]
\centering
\includegraphics[width=1\textwidth]{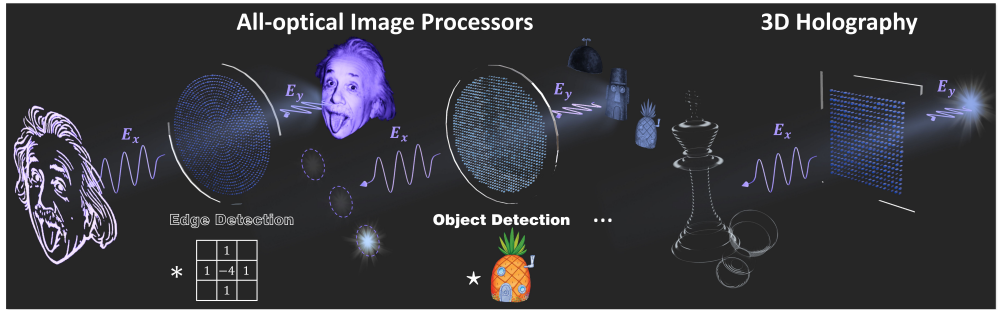}
\caption{Schematic of metasurface-based all-optical image processing and 3D holography, illustrating edge detection, object detection, and 3D hologram generation.}\label{fig 1}
\end{figure}

In this study, we introduce a metasurface-based platform for arbitrary all-optical image transformations, referred to as meta-operators. By integrating double-phase encoding and polarization multiplexing in metasurfaces, we experimentally realize multiple image processing functions, including one-dimensional (1D) and two-dimensional (2D) first-order differentiation, object detection via cross-correlation, vertex detection, and Laplace differentiation (Figure~\ref{fig 1}). These operations are implemented through linear and circular polarization multiplexing, enabling compact and versatile optical processing. Our approach is the first to experimentally achieve multiple image processing functions at visible wavelengths, highlighting the versatility of metasurfaces for computational optics. In addition to image processing, the polarization multiplexing strategy enables precise modulation of the complex amplitude of the light field at ultra-high resolution, making it well-suited for three-dimensional (3D) holography at visible wavelengths. As illustrated in Figure~\ref{fig 1}, we experimentally demonstrate high-fidelity volumetric holograms with clear stereoscopic reconstructions. These results highlight the potential of metasurfaces for analog image processing and compact, high-resolution holography, opening new opportunities for optical information processing and manipulation.

\section{Results}\label{sec2}

\subsection{Principle of meta-operators design}\label{subsec2_1}

\begin{figure}[H]
\centering
\includegraphics[width=1\textwidth]{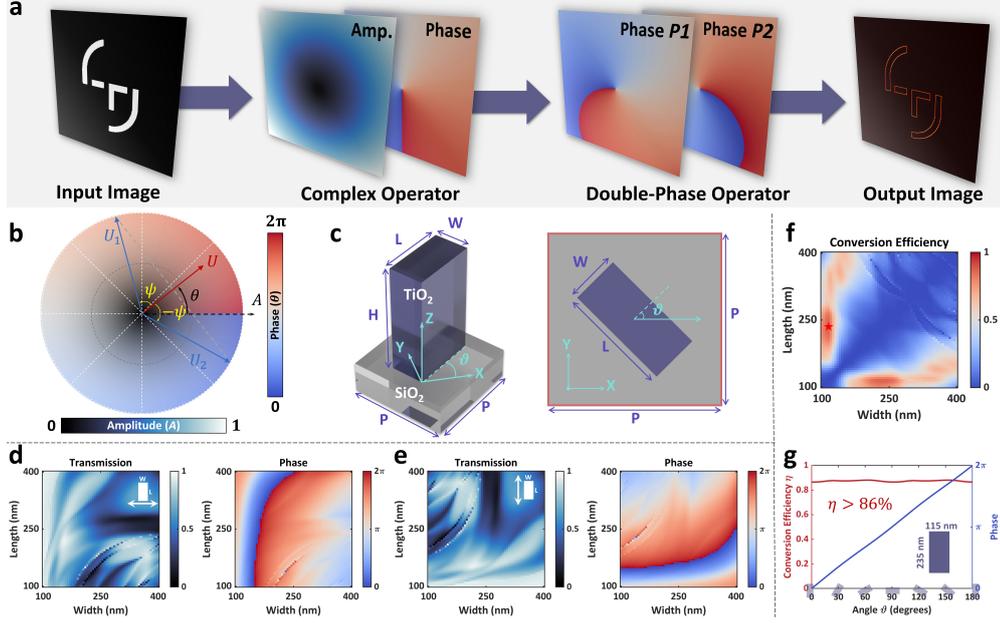}
\caption{Working principle and design of the meta-atoms. (a) Schematic of the double-phase encoded meta-operators for image processing based on polarization multiplexing. (b) Conceptual illustration of the complex-amplitude modulation strategy. (c) Schematic representation of TiO\textsubscript{2} rectangular nanopillar meta-atoms. (d) Transmission and phase shifts of meta-atoms as a function of width and length under x-polarized incidence. (e) Transmission and phase shifts under y-polarized incidence. (f) Polarization cross-conversion efficiency from LCP to RCP as a function of meta-atom width and length. (g) Polarization cross-conversion efficiency and geometric phase shifts of meta-atoms with a fixed length of 235 nm and width of 115 nm, plotted as a function of orientation angle from 0\textdegree{} to 180\textdegree{}}\label{fig 2}
\end{figure}

\noindent Image processing operations can be mathematically described as convolutional transformations, where an input image is convolved with a kernel function~\cite{gonzalez2006}. In optical systems, convolution naturally occurs in the Fourier domain, where a lens inherently acts as a Fourier transformer, mapping spatial features into the frequency domain~\cite{goodman2017introduction}. This enables direct spatial frequency manipulation, allowing convolutional operators to be embedded in the Fourier plane for real-time, parallel image processing. However, implementing arbitrary convolutional filters requires full complex-amplitude modulation—precise control over both amplitude and phase—which remains a key challenge, particularly in compact, high-resolution optical systems. To address this challenge, we implement meta-operators, which enable full complex-amplitude modulation in a compact and efficient form at visible wavelengths. These meta-operators process images by encoding the input light field with equal contributions in different polarization states. In the Fourier domain, the spatial spectrum of the input image is modulated by a complex filter acting as a computational operator. This transformation is realized through the combination of polarization multiplexing and double-phase encoding, which together enable precise optical transfer functions using a single passive metasurface. Figure~\ref{fig 2}(a)illustrates the working principle of the meta-operators. Meta-operator realize independent modulation of two orthogonal polarization components (\(P1\) and \(P2\)), encoding each with a distinct phase distribution. When these components interfere, they reconstruct the desired complex-amplitude modulation, forming a compact all-optical image processing platform.

In this framework, double-phase encoding~\cite{hsueh1978computer} is applied to decompose an arbitrary complex modulation \(U(x, y)\) by using two phase-only modulations, as illustrated in Figure~\ref{fig 2}(b). This representation is naturally expressed in polar coordinates, where the azimuthal coordinate represents the phase, ranging from \(0\) to \(2\pi\), while the radial coordinate represents the normalized amplitude, ranging from \(0\) to \(1\). A complex modulation \(U(x, y)\) with amplitude \(A(x, y)\) and phase \(\theta(x, y)\) can be decomposed into two equal-amplitude components \(U_1\) and \(U_2\), expressed as:
\begin{equation}
U(x, y) = A(x, y)e^{i\theta(x, y)} = B e^{i(\theta(x, y) + \psi(x, y))} + B e^{i(\theta(x, y) - \psi(x, y))},
\label{eq:eq1}
\end{equation}
\noindent where \( B = A_\text{max}/2 \) is a constant, and \( \psi(x, y) = \cos^{-1}\left[A(x, y)/A_\text{max}\right] \). This decomposition reformulates the complex modulation into two phase-only modulations, \( \theta(x, y) + \psi(x, y) \) and \( \theta(x, y) - \psi(x, y) \), enabling full complex-amplitude modulation using phase-only optics. To physically realize this phase-only encoding compactly and efficiently, we implement a single-layer metasurface composed of TiO\textsubscript{2} rectangular nanopillar meta-atoms,  as illustrated in Figure~\ref{fig 2}(c), which induce polarization-dependent phase shifts while maintaining unit transmission. The metasurface response is described by the Jones matrix \( \mathcal{M} \):
\begin{equation}
\mathcal{M}(x, y) = \mathbf{R}(\vartheta)
\begin{bmatrix}
e^{i\phi_x(x, y)} & 0 \\ 
0 & e^{i\phi_y(x, y)}
\end{bmatrix}
\mathbf{R}^{-1}(\vartheta),
\label{eq:eq2}
\end{equation}
\noindent where \(\mathbf{R}(\vartheta)\) represents the rotation matrix of a meta-atom with orientation angle \(\vartheta\), and \(\phi_x(x, y)\) and \(\phi_y(x, y)\) denote the phase shifts along the short and long axes of the meta-atoms, respectively. For visible-light operation at a wavelength of 532 nm, the metasurface is fabricated on a glass substrate using TiO\textsubscript{2} nanopillars with a fixed height (\(H = 600\) nm) and period (\(P = 450\) nm). The nanopillar width (\(W\)) and length (\(L\)) vary between 100 nm and 400 nm, enabling precise tuning of the local phase shift for orthogonal polarization states via birefringence. The transmission and phase shifts of the meta-atoms under x- and y-polarized incidence, obtained via finite element simulations, are shown in Figures~\ref{fig 2}(d,e). Further details are provided in Supplementary Information Section 5. In this work, we utilize both orthogonal linear and cross-circular polarization states to encode the required phase distributions. For experimental demonstration, the fabricated meta-operators consist of 4000 \(\times\) 4000 meta-atoms, corresponding to a physical size of 1.8 mm \(\times\) 1.8 mm.

For linear polarization multiplexing, the orientation angle of the meta-atoms \(\vartheta\) is set to zero, enabling independent phase control for x- and y-polarized light as \(\theta(x, y) + \psi(x, y)\) and \(\theta(x, y) - \psi(x, y)\). Under this configuration, the metasurface applies the following transformation:
\begin{equation}
\begin{bmatrix}
E_{x,\text{out}} \\
E_{y,\text{out}}
\end{bmatrix}
=
\begin{bmatrix}
e^{i\left(\theta(x,y) + \psi(x,y)\right)} & 0 \\
0 & e^{i\left(\theta(x,y) - \psi(x,y)\right)}
\end{bmatrix}
\begin{bmatrix}
E_{x,\text{in}} \\
E_{y,\text{in}}
\end{bmatrix}.
\label{eq:eq3}
\end{equation}
For an incident wave with equal x- and y-polarized components (\(E_{x,\text{in}} = E_{y,\text{in}} = E_0\)), the output fields (\(E_{x,\text{out}}\) and \(E_{y,\text{out}}\)) interfere, resulting in a combined output field:
\begin{equation}
E_{\text{interfere,out}} = \sqrt{2} E_0 \cos\psi(x,y) e^{i\theta(x,y)}.
\label{eq:eq4}
\end{equation}
This formulation demonstrates that linear polarization multiplexing enables complex-amplitude modulation with precise control over the transmitted wavefront.

As an alternative to linear polarization multiplexing, metasurfaces can be designed to operate with circular polarization states, enabling enhanced sensitivity to phase variations and making them particularly effective for specific computational operators. In this scheme, the meta-atoms function as half-wave plates with fixed geometric dimensions but varying orientation angles. According to Equation~\ref{eq:eq2}, when \(\phi_y - \phi_x = \pi\), an incident linearly polarized wave is converted into transmitted right- and left-circularly polarized (RCP and LCP) components, given by:
\begin{equation}
\begin{bmatrix}
E_{\text{RCP,out}} \\ 
E_{\text{LCP,out}}
\end{bmatrix}
=
\frac{1}{\sqrt{2}}
\begin{bmatrix}
1 & -i \\ 
1 & i
\end{bmatrix}
\begin{bmatrix}
E_{x,\text{out}} \\ 
E_{y,\text{out}}
\end{bmatrix}
=
\frac{1}{\sqrt{2}}
\begin{bmatrix}
e^{i(\phi_x + 2\vartheta)} \\ 
e^{i(\phi_x - 2\vartheta)}
\end{bmatrix}.
\label{eq:eq5}
\end{equation}
\noindent As a result, the RCP and LCP components acquire opposite phase delays, a phenomenon known as the Pancharatnam-Berry (PB) phase~\cite{arbabi2015dielectric,balthasar2017metasurface}. Leveraging this geometric phase modulation, metasurfaces enable a variety of computational optical operations, including second-order derivatives (\( \frac{\partial^2}{\partial x \partial y} \), \( \nabla^2 \)) and integral transformations that exhibit even-symmetric or parity-invariant properties, such as Gaussian smoothing and the Hilbert transform. To effectively realize PB-phase encoding, the geometric parameters of the meta-atoms must be carefully optimized to ensure high efficiency and precise phase modulation. We evaluated the modulation efficiency of TiO\textsubscript{2} nanopillars based on Figures~\ref{fig 2}(d,e). The results, shown in Figure~\ref{fig 2}(f), indicate that nanopillars with a length of 235 nm, width of 115 nm, and height of 600 nm (highlighted by the red pentagram) exhibit high efficiency. The corresponding phase shift and efficiency as a function of orientation angle, presented in Figure~\ref{fig 2}(g), confirm the suitability of this metasurface design for implementing complex optical transfer functions. Further details are provided in Supplementary Information Section 6.

\subsection{First-order differentiation meta-operators via linear polarization multiplexing}\label{subsec2_2}

\begin{figure}[H]
\centering
\includegraphics[width=1\textwidth]{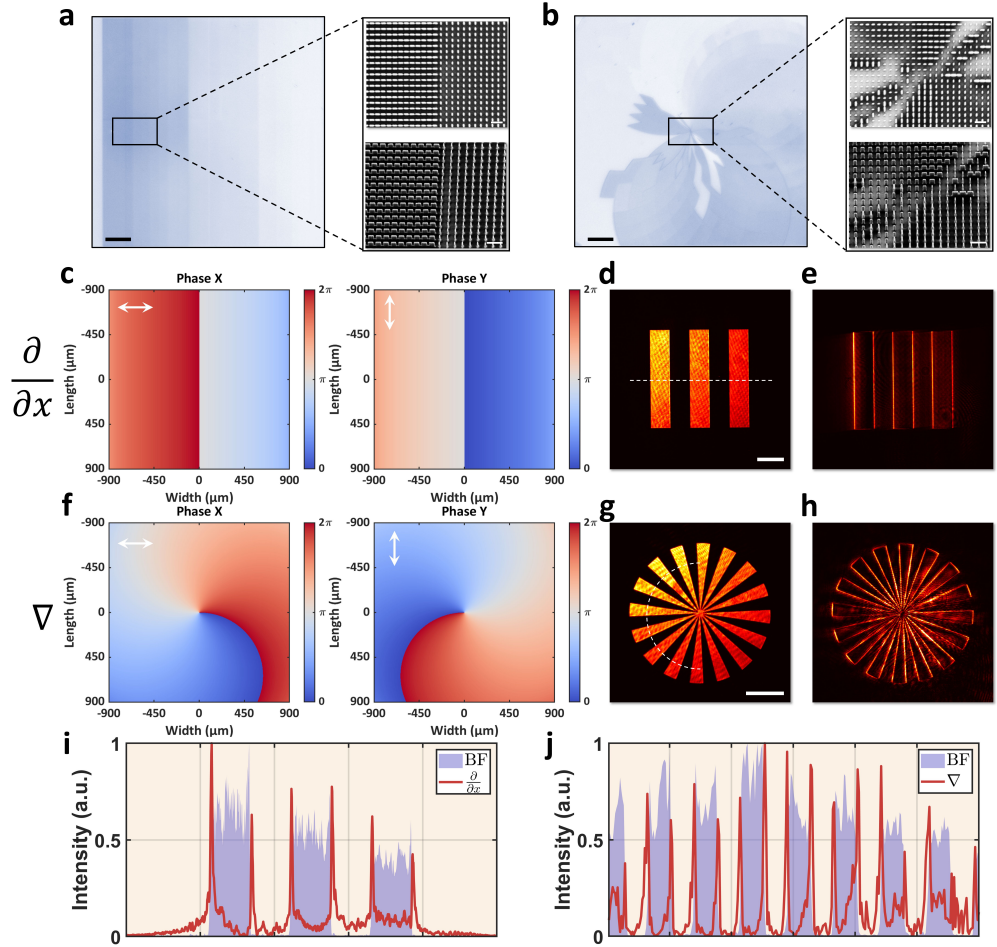}
\caption{First-order differentiation meta-operators via linear polarization multiplexing. (a) Optical and scanning electron microscopy (SEM) images of the fabricated meta-operator for 1D first-order differentiation. (b) Optical and SEM images of the fabricated meta-operator for 2D first-order differentiation. (c) Phase maps of the 1D first-order differentiation meta-operator under x- and y-polarized illumination. (d, e) Bar pattern images before and after processing with the 1D first-order differentiation meta-operator. (f) Phase maps of the 2D first-order differentiation meta-operator under x- and y-polarized illumination. (g, h) Spoke pattern images before and after processing with the 2D first-order differentiation meta-operator. (i, j) Cross-sectional intensity profiles corresponding to (d, e) and (g, h), respectively. Scale bars: optical images, 200 \textmu m; SEM images, 1 \textmu m; experimental images (e, h), 200 \textmu m.}\label{fig 3}
\end{figure}

We first demonstrate the capability of meta-operators by implementing first-order differentiation in both 1D and 2D cases. First-order differentiation plays a fundamental role in image processing, particularly in edge detection and feature extraction, which enhance structural details and are essential for applications such as image segmentation, object detection, and texture analysis.

The design of these meta-operators is based on the Fourier representation of first-order differentiation. The Fourier transform of a function \( f(x) \) is given by:
\begin{equation}
\hat{f}(k) = \int_{-\infty}^{\infty} f(x) e^{-2\pi i kx} dx.
\label{eq:eq6}
\end{equation}
Applying the Fourier transform to the first-order derivatives of \( f(x, y) \) yields:
\begin{equation}
\mathcal{F} \left( \frac{\partial}{\partial x} f(x,y) \right) = (2\pi i k_x) \hat{f}(k_x, k_y), \quad
\mathcal{F} \left( \frac{\partial}{\partial y} f(x,y) \right) = (2\pi i k_y) \hat{f}(k_x, k_y).
\label{eq:eq7}
\end{equation}
This result shows that first-order differentiation in the spatial domain translates to multiplication by \( 2\pi i k_x \) or \( 2\pi i k_y \) in the Fourier domain. To extend this to two dimensions in a compact form, we introduce the complex coordinate:
\begin{equation}
\xi = x + i y.
\label{eq:eq8}
\end{equation}
With this notation, differentiation can be expressed as a single complex operator:
\begin{equation}
\frac{\partial}{\partial \xi} = \frac{\partial}{\partial x} + i \frac{\partial}{\partial y}.
\label{eq:eq9}
\end{equation}
Applying the Fourier transform to this expression gives:
\begin{equation}
\mathcal{F} \left( \frac{\partial}{\partial \xi} f(x,y) \right) = 2\pi i (k_x + i k_y) \hat{f}(k_x, k_y).
\label{eq:eq10}
\end{equation}
Based on Equations~\ref{eq:eq7} and \ref{eq:eq10}, we design meta-operators for 1D first-order differentiation along the x-direction and 2D differentiation. The corresponding amplitude and phase modulations required in the Fourier domain are shown in Figure S2(a, b). These meta-operators are physically realized using double-phase encoding with linear polarization multiplexing.

For 1D first-order differentiation along the x-direction, the designed meta-operator is shown in Figure~\ref{fig 3}(a), with the corresponding phase maps for x- and y-polarized light depicted in Figure~\ref{fig 3}(c). A bar pattern image (Figure S16(a)) serves as the test input. The experimental setup, illustrated in Figure S13(a), captures the image under bright-field (BF) illumination, as shown in Figure~\ref{fig 3}(d). Upon propagation through the meta-operator, the transformed image in Figure~\ref{fig 3}(e) exhibits strong lateral edge enhancement, characterized by high-contrast transitions at intensity boundaries. This differentiation effect is quantitatively verified by the cross-sectional intensity profile in Figure~\ref{fig 3}(i), which reveals distinct intensity peaks at the edges, in agreement with the theoretical predictions shown in Figure S2(f).

For the 2D first-order differentiation meta-operator, shown in Figure~\ref{fig 3}(b), the phase maps for x- and y-polarized light are depicted in Figure~\ref{fig 3}(f). A spoke pattern image (Figure S16(b)) serves as the test input, with the captured image under bright-field illumination shown in Figure~\ref{fig 3}(g). The processed image in Figure~\ref{fig 3}(h) demonstrates edge enhancement across all directions. This effect is further quantified by the cross-sectional intensity profiles in Figure~\ref{fig 3}(j), which show distinct intensity variations corresponding to edge locations. The experimental results show strong agreement with theoretical predictions, as shown in Figure S2(g), confirming the accuracy of the designed meta-operator.

The demonstrated first-order differentiation meta-operators provide a highly compact and passive approach to optical edge detection, making them well-suited for real-time applications in machine vision, biomedical diagnostics, and computational optics. Their ability to extract structural features directly from optical fields without computational post-processing offers significant advantages for ultrafast imaging systems, high-throughput microscopy, and autonomous navigation, where real-time processing and energy efficiency are crucial.

\subsection{Cross-correlation meta-operators via linear polarization multiplexing}\label{subsec2_3}

\begin{figure}[H]
\centering
\includegraphics[width=1\textwidth]{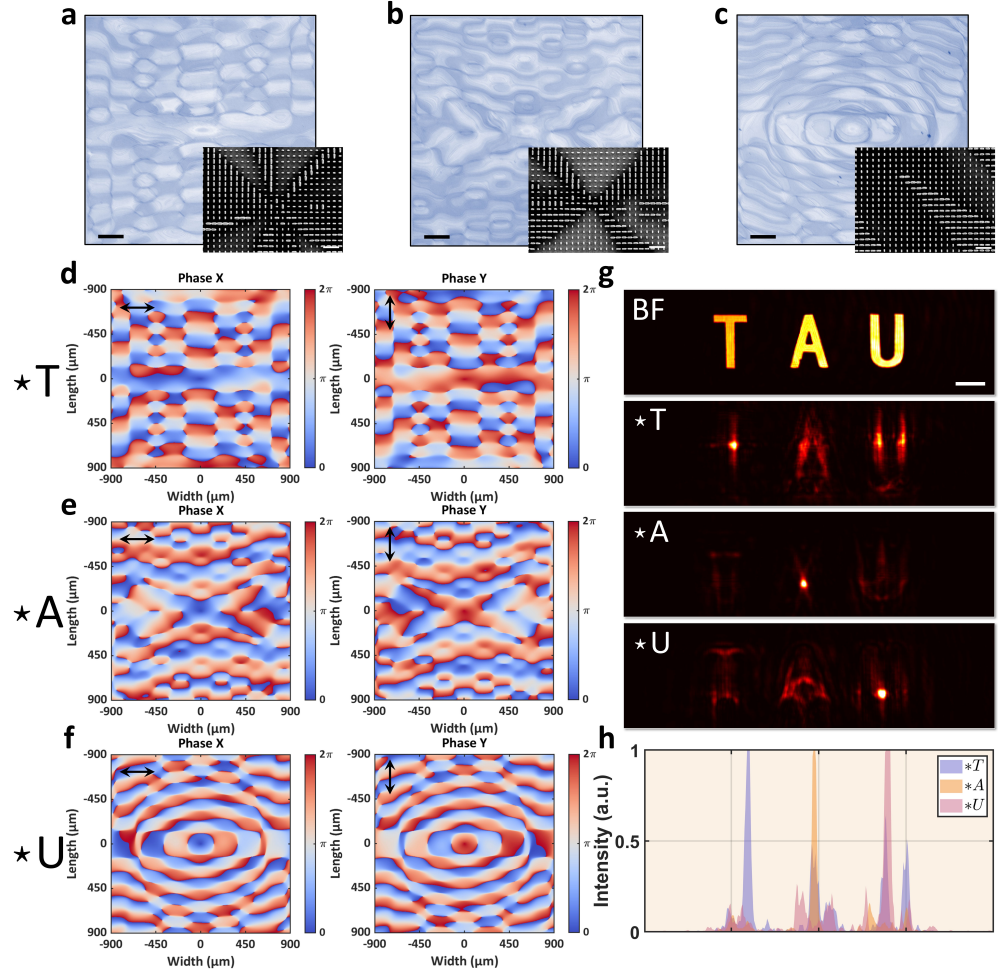}
\caption{Cross-correlation meta-operators via linear polarization multiplexing. (a–c) Optical and SEM images of the fabricated meta-operators designed for detecting the letters \textit{T}, \textit{A}, and \textit{U}, respectively. (d–f) Phase maps of the corresponding cross-correlation meta-operators under x- and y-polarized light for the letters \textit{T}, \textit{A}, and \textit{U}, respectively. (g) Input \textit{TAU} images before and after processing with the letter detection meta-operators. (h) Cross-sectional intensity profiles corresponding to (g). Scale bars: optical images, 200 \textmu m; SEM images, 1 \textmu m; experimental images (g, h), 200 \textmu m.}\label{fig 4}
\end{figure}

In addition to edge detection via first-order differentiation, meta-operators can also perform more advanced image processing tasks. One such task is cross-correlation, which identifies specific patterns within an image by measuring their similarity to a reference template. Unlike differentiation, which enhances structural features by highlighting intensity variations, cross-correlation enables object recognition and alignment by directly matching image patterns. Here, we implement all-optical cross-correlation using meta-operators designed with linear polarization multiplexing.

Mathematically, cross-correlation is equivalent to convolution with a flipped target function:
\begin{equation}
(I \star T)(x,y) = (I * T^F)(x,y),
\label{eq:eq11}
\end{equation}
\noindent where \( T^F(x,y) = T(-x,-y) \) denotes the flipped target function~\cite{gonzalez2006}. Applying the Fourier transform to both sides and using the convolution theorem, we obtain:
\begin{equation}
\mathcal{F} (I \star T) = \mathcal{F} (I * T^F) = \mathcal{F} (I) \cdot \mathcal{F} (T^F).
\label{eq:eq12}
\end{equation}
Since spatial flipping in real space corresponds to complex conjugation in the Fourier domain, this simplifies to:
\begin{equation}
\mathcal{F} (I \star T) = \mathcal{F} (I) \cdot \overline{\mathcal{F} (T)}.
\label{eq:eq13}
\end{equation}
\noindent To realize cross-correlation meta-operators for arbitrary patterns, the target patterns are first flipped and Fourier transformed to generate the corresponding complex transfer functions. These are then mapped onto phase distributions for x- and y-polarized light, enabling implementation through meta-operators based on linear polarization multiplexing.

As a proof of concept, we demonstrate cross-correlation meta-operators for detecting specific letters within an image containing the word \textit{TAU}, as shown in Figure S16(c). The inset images of the letters \textit{T}, \textit{A}, and \textit{U} are separately used to design the corresponding meta-operators. The designed meta-operators selectively enhance the target letters, making them appear as distinct bright intensity spots at their respective locations after processing. The corresponding complex transfer functions for these letters, which define the filtering process, are shown in Figure S2(c–e). The fabricated cross-correlation meta-operators are shown in Figure~\ref{fig 4}(a–c), with their respective phase distributions under x- and y-polarized light in Figure~\ref{fig 4}(d–f). The experimental setup, depicted in Figure S13(b), was used for validation. The input image under bright-field illumination, along with the processed results from the cross-correlation meta-operators, is shown in Figure~\ref{fig 4}(g). Strong intensity spots appear at the target letter positions, demonstrating the effectiveness of the designed meta-operators for optical pattern recognition. The corresponding theoretical predictions are presented in Figure S2(h–j), demonstrating consistency with the experimental results.

These cross-correlation meta-operators provide a highly compact and passive approach to optical pattern recognition, enabling rapid and real-time identification of target features. Unlike conventional methods that rely on digital computation, these meta-operators execute correlation operations at the speed of light, eliminating latency and power consumption associated with electronic processing. This makes them particularly advantageous for applications requiring instantaneous decision-making, such as autonomous vision systems, secure optical authentication, and real-time optical tracking.

\subsection{Second-order differentiation meta-operators via circular polarization multiplexing}\label{subsec2_4}

\begin{figure}[H]
\centering
\includegraphics[width=1\textwidth]{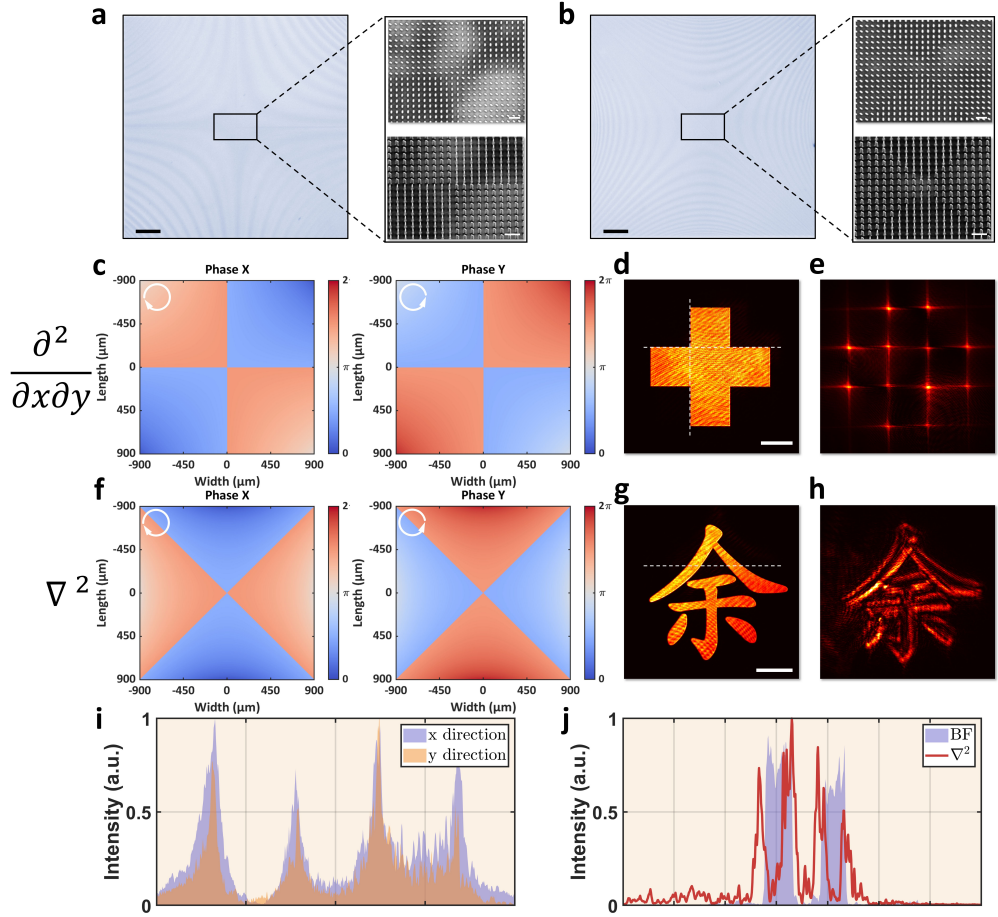}
\caption{Second-order differentiation meta-operators via circular polarization multiplexing. (a) Optical and SEM images of the fabricated meta-operator for vertex detection. (b) Optical and SEM images of the fabricated meta-operator for Laplacian differentiation. (c) Phase maps of the vertex detection meta-operator under RCP and LCP illumination. (d, e) Cross-pattern image before and after processing with the vertex detection meta-operator. (f) Phase maps of the Laplacian differentiation meta-operator under RCP and LCP illumination. (g, h) Image of the Chinese character ‘YU’ before and after processing with the Laplacian meta-operator. (i, j) Cross-sectional intensity profiles corresponding to (e) and (h), respectively. Scale bars: optical images, 200 \textmu m; SEM images, 1 \textmu m; experimental images (e, h), 200 \textmu m.}\label{fig 5}
\end{figure}

To enhance the versatility of meta-operators, we implement circular polarization multiplexing, enabling higher computational precision. As a demonstration, we realize second-order optical differentiation, which extends beyond first-order differentiation by detecting higher-order structural variations and curvature changes. Specifically, we design meta-operators for vertex detection and Laplacian differentiation, showcasing their potential for high-precision image analysis and optical computing.

For vertex detection, which extracts second-order mixed derivatives, the corresponding transfer function is defined as:
\begin{equation}
\mathcal{F} \left( \frac{\partial^2}{\partial x \partial y} \right) = - (2\pi k_x)(2\pi k_y).
\label{eq:eq14}
\end{equation}
This function selectively enhances frequency components corresponding to intensity variations along both the x- and y-directions, making it particularly effective for identifying structural junctions and intersecting features. In contrast, the Laplacian operator, which computes the sum of second-order derivatives in both spatial dimensions, functions as an isotropic high-pass filter:
\begin{equation}
\mathcal{F} \left( \nabla^2 f(x,y) \right) = - \left[ (2\pi k_x)^2 + (2\pi k_y)^2 \right] \hat{f}(k_x, k_y).
\label{eq:eq15}
\end{equation}
Unlike vertex detection, which emphasizes directional intensity variations, the Laplacian operator enhances fine details and high-frequency features in an isotropic manner, making it effective for edge sharpening and contrast enhancement. To validate these operations, we implement the designed transfer functions using circular polarization multiplexing.

For vertex detection, the fabricated meta-operator is presented in Figure~\ref{fig 5}(a), with its phase maps for RCP and LCP illumination shown in Figure~\ref{fig 5}(c). A cross-pattern image (Figure S16(d)) is used as the input, and the captured bright-field image (experimental setup in Figure S13(a)) is depicted in Figure~\ref{fig 5}(d). The processed output in Figure~\ref{fig 5}(e) clearly enhances the pattern’s corners. The cross-sectional intensity profiles along the x- and y-directions (Figure~\ref{fig 5}(i)) further validate effective corner enhancement, consistent with theoretical predictions in Figure S3(c).

For Laplacian differentiation, the fabricated meta-operator is shown in Figure~\ref{fig 5}(b), with the corresponding phase maps for RCP and LCP illumination in Figure~\ref{fig 5}(f). An image of the Chinese character \textit{YU} (Figure S16(e)) serves as the input. The captured bright-field image is shown in Figure\ref{fig 5}(g), while the processed result in Figure~\ref{fig 5}(h) reveals pronounced edge sharpening and fine structural detail enhancement. The cross-sectional intensity profiles in Figure~\ref{fig 5}(j) confirm these effects, demonstrating strong high-frequency feature extraction. The theoretical result, provided in Figure S3(d), shows excellent agreement with experimental observations.

These results demonstrate the unique advantages of circular polarization multiplexing in extracting higher-order structural features. Second-order differentiation meta-operators enable precise feature localization and structural variation detection, making them well-suited for applications such as defect inspection in semiconductor lithography, quantitative phase imaging for label-free biomedical analysis, and high-resolution feature extraction in surface metrology.

\subsection{3D meta-hologram}\label{subsec2_5}

\begin{figure}[H]
\centering
\includegraphics[width=1\textwidth]{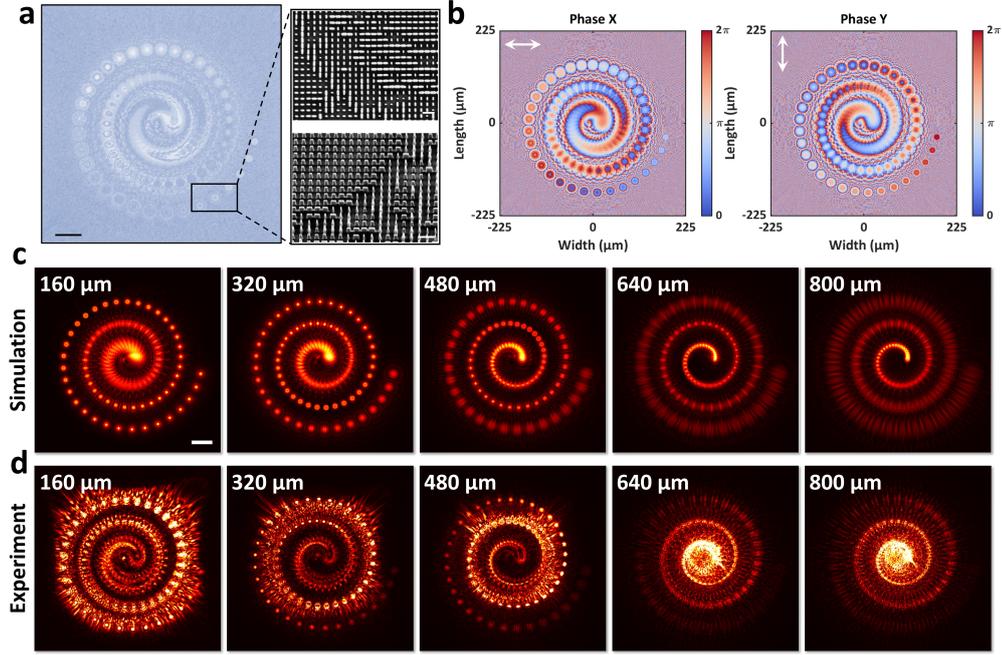}
\caption{3D meta-hologram via linear polarization multiplexing. (a) Optical and SEM images of the fabricated spiral-pattern 3D meta-hologram. (b) Phase distributions of the meta-hologram for x- and y-polarized illumination. (c, d) Simulated and experimentally measured intensity distributions of the reconstructed 3D hologram at different focal planes. Scale bars: optical images, 50 \textmu m; SEM images, 1 \textmu m; hologram images, 50 \textmu m.}\label{fig 6}
\end{figure}

Having demonstrated the versatility of meta-operators for analog image processing, we extend this platform to holographic wavefront shaping. Specifically, we design and fabricate metasurface-based holograms (meta-holograms) that utilize the same polarization-multiplexed encoding strategy to achieve high-fidelity 3D optical reconstructions~\cite{jiang2019metasurface,ni2013metasurface}. As a proof of concept, we implement a 3D meta-hologram that reconstructs a spiral dot array pattern, where dots are uniformly distributed from 0 \textmu m to 800 \textmu m in depth at 8 \textmu m intervals, as illustrated in Figure S6(a). The meta-hologram, consisting of 1000 \(\times\) 1000 meta-atoms (450 \textmu m \(\times\) 450 \textmu m), is designed by numerically back-propagating the target 3D light field to the meta-hologram plane (see Supplementary Information Section 3 for details).

The fabricated meta-hologram is characterized using the experimental setup depicted in Figure S13(c). Optical and SEM images of the meta-hologram are shown in Figure~\ref{fig 6}(a). The encoded phase distributions for x- and y-polarized light, computed via double-phase encoding as described by Equation~\ref{eq:eq1}, are presented in Figure~\ref{fig 6}(b). The theoretically reconstructed 3D intensity distributions at different focal depths are shown in Figure~\ref{fig 6}(c), with the corresponding experimental results in Figure~\ref{fig 6}(d), demonstrating excellent agreement. Additional 3D hologram demonstrations and the recorded light field intensity variations across different depths are provided in Supplementary Information Section 4 and the accompanying videos.

These results demonstrate the effectiveness of polarization-multiplexed encoding for high-fidelity 3D holography. By enabling precise polarization-controlled depth modulation, this approach achieves subwavelength modulation resolution down to 450 nm, ensuring accurate volumetric reconstructions. Further optimization of the meta-atom design could enable even finer resolution. The ability to tailor holographic wavefronts with such precision makes meta-holograms particularly promising for applications requiring fine depth control, such as dynamic volumetric displays~\cite{blanche2021holography,liu2023metasurface}, high-density optical data storage~\cite{hesselink2004holographic,curtis2011holographic}, and advanced holographic encryption~\cite{zhao2018multichannel,qu2020reprogrammable}.

\section{Discussion}\label{sec3}

\noindent In summary, we have developed and experimentally demonstrated meta-operators, a compact and scalable platform that enables all-optical image processing and extends to holographic wavefront shaping. By leveraging double-phase encoding and polarization multiplexing, we realized first- and second-order differentiation, cross-correlation, and volumetric holography, achieving precise optical transformations in a passive, single-layer metasurface. These results establish a versatile approach for encoding complex optical transfer functions, providing a practical solution for real-time analog image processing and compact, high-resolution holography. As metasurface technology advances, this platform offers a promising pathway toward ultrafast, energy-efficient optical computing, next-generation computational microscopy, and intelligent sensing, expanding the possibilities of future photonic information processing.

\section{Methods}\label{sec4}

\subsection{Simulation}\label{subsec4_1} The transmission and phase shift of each metasurface unit cell were simulated using the RF Module in COMSOL Multiphysics. The refractive index of TiO\textsubscript{2} was measured via ellipsometry and determined to be 2.47 with no extinction, while the refractive index of the glass substrate was set to 1.53. Light propagation in free space was modeled using the angular spectrum method~\cite{goodman2017introduction,matsushima2009band}. All numerical simulations, including meta-operator and meta-hologram designs, were performed in MATLAB.

\subsection{Fabrication}\label{subsec4_2} The metasurface was fabricated by first depositing a 600 nm thick TiO\textsubscript{2} film onto glass substrates, followed by spin-coating an electron-beam resist. A pre-designed 2D mask pattern was transferred to the resist via electron-beam lithography. After patterning, a chromium hard mask was deposited and employed for etching the TiO\textsubscript{2} layer using reactive ion etching. Finally, the chromium layer was removed, revealing the metasurface structure. Additional fabrication details are provided in Supplementary Information Section 8.

\subsection{Experimental setup}\label{subsec4_3} 
To evaluate the performance of the meta-operators, input images were illuminated with equal contributions in different polarization states using a polarizer and then relayed through a 4\(f\) system, where the meta-operators were placed at the Fourier plane. A second polarizer, positioned before the image sensor, ensured interference between different polarization components, allowing the processed images to be captured. For 3D meta-holograms, the meta-holograms were illuminated with a collimated laser beam containing equal amounts of orthogonal linear polarization states. After modulation, the reconstructed holographic intensity distributions at different depths were recorded by an image sensor using an objective and a tube lens. A second polarizer ensured the interference of modulated polarization components before detection. Further details are provided in Supplementary Information Section 7.

\backmatter
\bmhead{Dat availability}
The data supporting this study are not publicly available but can be provided by the corresponding author upon reasonable request.
\bmhead{Supplementary information}
Supplementary information is available.

\bmhead{Acknowledgements}
L.Y. and H.C. acknowledge financial support from the European Union’s Horizon 2020 research and innovation programme under the Marie Skłodowska-Curie grant agreement No 956770.

\bibliography{sn-bibliography}

\end{document}